%% file: main.tex
\documentclass[a4paper]{article}
\usepackage{fullpage}
\usepackage{xspace}
\usepackage{subcaption}							
\usepackage{latexsym}
\usepackage{graphicx}
\usepackage{mathptmx}
\usepackage{wrapfig}
\usepackage{amsmath}
\usepackage{amsfonts}
\usepackage{amssymb}
\usepackage{amsbsy}
\usepackage{amsthm}

\newcommand{\SGREEN}{\textsc{green}\xspace}
\newcommand{\SYELLOW}{\textsc{yellow}\xspace}
\newcommand{\SRED}{\textsc{red}\xspace}
\newcommand{\SMANUAL}{\textsc{manual}\xspace}
\newcommand{\STOMANUAL}{\textsc{going\_manual}\xspace}
\newcommand{\STOAUTO}{\textsc{going\_auto}\xspace}
\newcommand{\SBREAK}{\textsc{break}\xspace}
\newcommand{\SWORKING}{\textsc{working}\xspace}

\newcommand{\EGREEN}{\textit{show\_green}\xspace}
\newcommand{\EYELLOW}{\textit{show\_yellow}\xspace}
\newcommand{\ERED}{\textit{show\_red}\xspace}
\newcommand{\EMANUAL}{\textit{toManual}\xspace}
\newcommand{\EAUTO}{\textit{toAuto}\xspace}

\newcommand{\EOFF}{\textit{turn\_off}\xspace}

\newcommand{\DELAYGREEN}{\textit{$57$}\xspace}
\newcommand{\DELAYYELLOW}{\textit{$3$}\xspace}
\newcommand{\DELAYRED}{\textit{$60$}\xspace}
\newcommand{\DELAYBREAK}{\textit{$120$}\xspace}
\newcommand{\DELAYWORKING}{\textit{$360$}\xspace}

\usepackage[pdftex,colorlinks=true,urlcolor=blue,citecolor=black,anchorcolor=black,linkcolor=black]{hyperref}

\begin{document}

\title{Extending the DEVS Formalism with Initialization Information}

    \newcommand{\footremember}[2]{%
        \footnote{#2}
        \newcounter{#1}
        \setcounter{#1}{\value{footnote}}%
    }
    \newcommand{\footrecall}[1]{%
        \footnotemark[\value{#1}]%
    }

\author{
    \begin{tabular}{cc}
        Yentl Van Tendeloo\footremember{UA}{University of Antwerp, Belgium}  &   Hans Vangheluwe\footrecall{UA} \footnote{Flanders Make, Belgium} \footnote{McGill University, Montr\'{e}al, Canada}\\
    \end{tabular} \\
    \{Yentl.VanTendeloo,Hans.Vangheluwe\}@uantwerpen.be
}

\date{}

\maketitle

\input{0-abstract}


\input{1-introduction}
\input{2-devs}
\input{3-initial_total_state}
\input{5-related_work}
\input{6-conclusion}

\bibliographystyle{plain}
\bibliography{bibliography}

\section*{Acknowledgements}
This work was partly funded by a PhD fellowship from the Research Foundation - Flanders (FWO).
This research was also partially supported by Flanders Make vzw, the Flemish strategic research centre for the manufacturing industry.
The authors wish to thank Joey De Pauw for pointing out an inconsistency between $q_{init}$ and the specification of $\delta_{int}$ in our lecture notes.

\end{document}

%% file: 0-abstract.tex
DEVS is a popular formalism to model system behaviour using a discrete-event abstraction.
The main advantages of DEVS are its rigourous and precise specification, as well as its support for modular, hierarchical construction of models.
DEVS frequently serves as a simulation ``assembly language'' to which models in other formalisms are translated, either giving meaning to new (domain-specific) languages, or reproducing semantics of existing languages.
Despite this rigourous definition of its syntax and semantics, initialization of DEVS models is left unspecified in both the Classic and Parallel DEVS formalism definition.
In this paper, we extend the DEVS formalism by including an initial total state.
Extensions to syntax as well as denotational (closure under coupling) and 
operational semantics (abstract simulator) are presented.
The extension is applicable to both main variants of the DEVS formalism.
Our extension is such that it adds to, but does not alter the original specification.
All changes are illustrated by means of a traffic light example.

%% file: 1-introduction.tex
\section{Introduction}
\label{sec:introduction}
DEVS~\cite{ClassicDEVS} is a popular formalism to model system behaviour using a 
discrete-event abstraction.
With this abstraction, only a finite number of pertinent events can occur during 
any bounded time interval.
In reaction to events, the model's state variable values change instantaneously.
The state remains unaltered between event occurrences.
The main advantages of DEVS are its rigourous and precise specification, 
as well as its support for modular, hierarchical model construction.
DEVS frequently serves as a simulation ``assembly language'' to which models in 
other formalisms are translated, either giving meaning to new (domain-specific) languages, 
or reproducing semantics of existing languages~\cite{DEVSbase}.
Models in different formalisms can hence be meaningfully combined by mapping them onto DEVS.

Despite the rigour of its syntax and semantics, the initialization of models is unspecified in both Classic~\cite{ClassicDEVS} and Parallel DEVS~\cite{ParallelDEVS,ParallelDEVSAbsSim}.
Initialization is important for several reasons.
First, it allows for unaltered reuse of models, for example from a model library.
A user should not have to know about, let alone modify, the various internal states 
and configurations of a reused model.
Second, initialization is required when restarting a simulation run after it was interrupted, 
e.g., for fault tolerance reasons. The state has to be 
re-initialized to the last know state, from which simulation can subsequently resume 
as if it was never interrupted.
Third, initialization is required for dynamic structure and hybrid systems, 
where the simulation is interrupted due to some system condition, the model is altered, 
re-initialized to a consistent state, and the simulation resumed.

As it is not part of the specification of the DEVS formalism,
different implementations have widely varying ways of supporting initialization, 
impeding model reuse across simulator implementations.
Additionally, realistic initial conditions can often not be expressed without 
adding artificial initialization behaviour to a model.
In this paper, a traffic light example is used to illustrate the need for 
adding initialization to the DEVS formalism.
The effects of the extension on both denotational (closure under coupling, or flattening) 
and operational semantics (the abstract simulator) are described and 
demonstrated using the running example.
Note that this extension only specifies initialization, leaving all other 
aspects of DEVS untouched.
Our extension is implemented in the PythonPDEVS~\cite{JDF} simulation tool, 
which is capable of simulating both Classic and Parallel DEVS.
As the addition to both is similar, only Classic DEVS is described in detail in this paper.

The remainder of this section presents our motivating example at a conceptual level.
Section~\ref{sec:devs} briefly recaps the Classic DEVS formalisms, without initialization, 
and models the example in it.
Section~\ref{sec:total_state} presents our extension to the formalism, 
the initial total state, and describes its influence on the specification 
and both of its semantics definitions.
Changes are demonstrated on the running example.
Section~\ref{sec:related_work} presents related work, mainly referring to current 
methods of initialization and tool support.
Section~\ref{sec:conclusion} concludes the paper.

\subsection{Motivating Example}
\begin{figure}
    \center
    \includegraphics[width=0.8\textwidth]{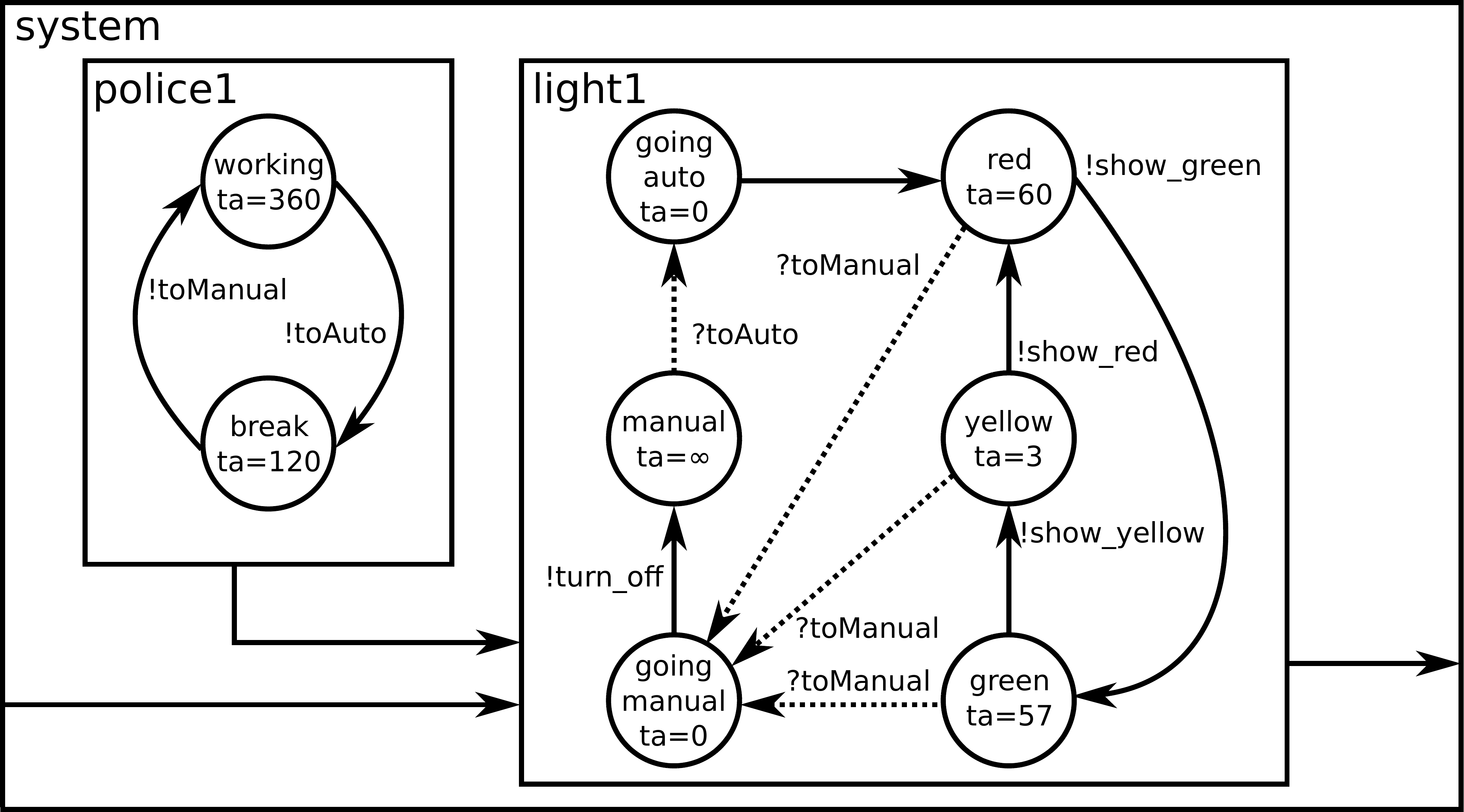}
    \caption{Visual representation of the traffic light model.}
    \label{fig:visual_model}
\end{figure}

To illustrate the lack of initialization, and the ensuing problems, a minimal example 
is used throughout this paper.
This running example consists of a simple traffic light which can run autonomously 
(i.e., alternate between red, yellow, and green) or be interrupted (i.e., be turned off).
This traffic light interacts with a policeman, who alternates between ``working'' 
(while the traffic light is turned off) and ``taking break'' 
(while the traffic light is working autonomously).
For the sake of readability of the behaviour traces, 
artificifial timings were chosen in the model shown in Figure~\ref{fig:visual_model}.
A full Classic DEVS specification is given in Section~\ref{sec:devs}.

From the visual representation of the model, there is no indication in which state 
a simulation starts.
For example, should the traffic light start autonomously or should it be turned off?
And what about the policeman?
This is the first problem: there is no initial state specified in the DEVS formalism.

\begin{figure}
    \center
    \begin{minipage}{0.55\textwidth}
        \includegraphics[width=\textwidth]{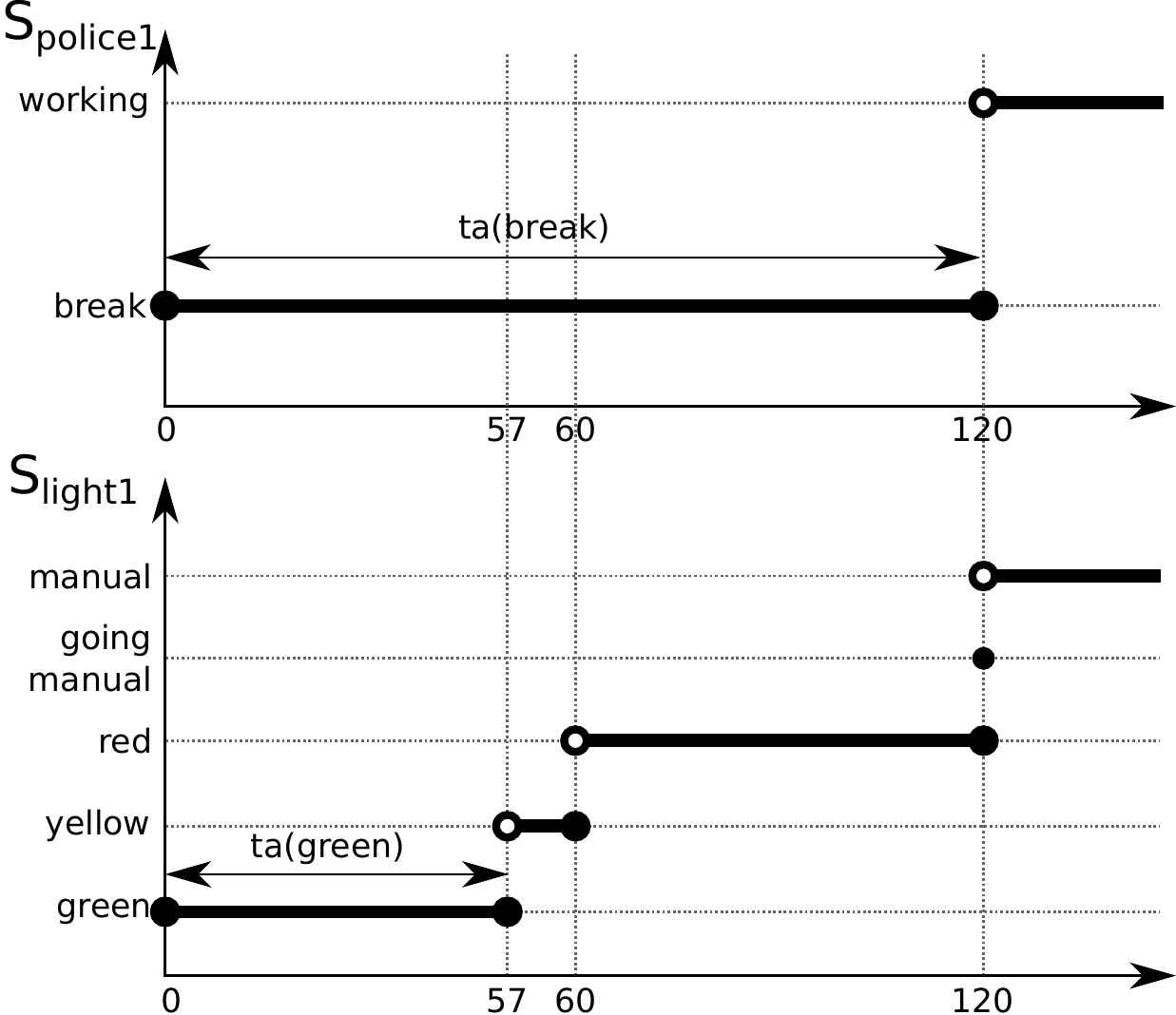}
        \caption{Initialized model without specified initial elapsed times for the component models.}
        \label{fig:trace_elapsed_0}
    \end{minipage}
    \hfill
    \begin{minipage}{0.4\textwidth}
        \includegraphics[width=\textwidth]{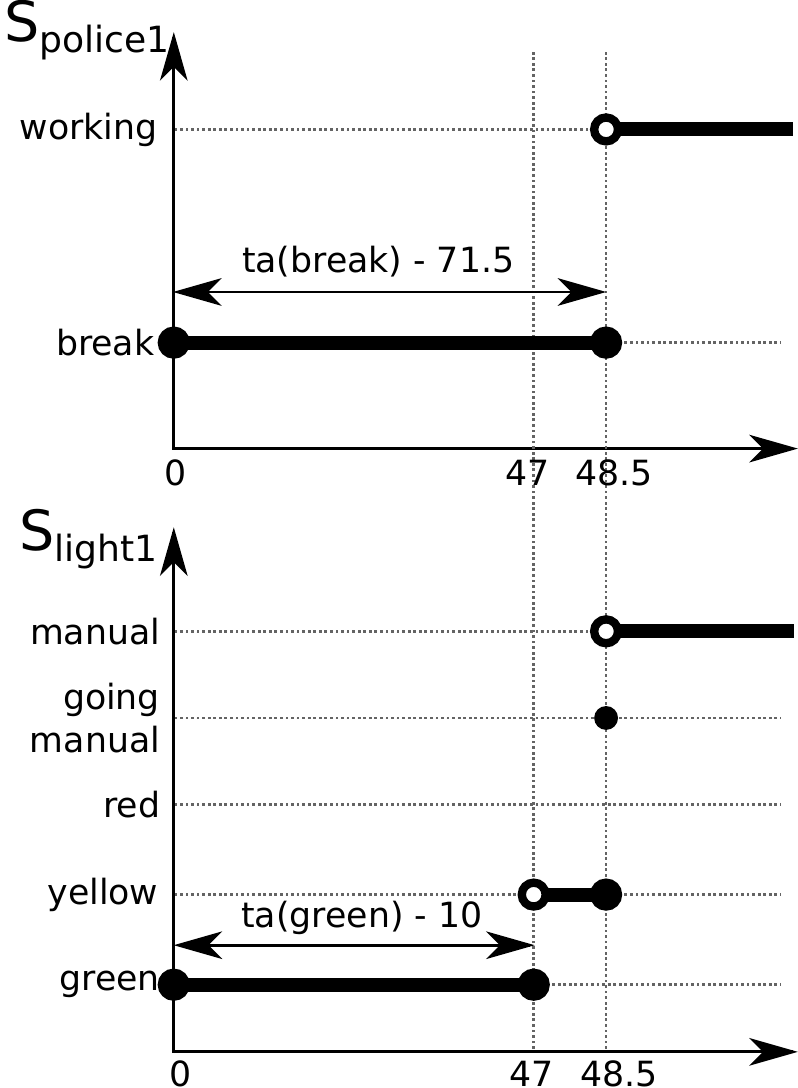}
	    \caption{Initialized model with specified initial elapsed times: 
	    start of component models' behaviours are shifted in time.}
        \label{fig:trace_elapsed_changed}
    \end{minipage}
\end{figure}

Assuming that some initial sequential state is given, a simulation can be performed, 
leading to the behaviour traces shown in 
Figure~\ref{fig:trace_elapsed_0}\footnote{For the sake of readability, none of the figures are to scale.}.
There is however no flexibility in when the policeman can trigger an event.
Without altering the delays in the policeman model, it is impossible for the policeman 
to interrupt the trafficlight at an arbitrary point in time: it must be at time \DELAYBREAK 
or \DELAYWORKING, as these are the time delays in the policaman model.
In both cases, however, this coincides with the time at which the light changes 
from red to green:
it is impossible to interrupt at any other time with this model.
To allow for arbitrary times, the model needs to be modified, adding artificial states, 
thereby introducing accidental and unnecessary complexity.
This is an indication that the formalism in its current form lacks expressiveness 
(or in this case, is under-specified).
Furthermore, the model remains incompletely initialized for as long as even a single 
atomic model is still in its initialization state.
Ideally, it should be possible to shift behaviour traces of component models independently
with respect to one another, such that the same models can be reused unchanged, 
but that for example the policeman starts in between two state changes, 
as shown in Figure~\ref{fig:trace_elapsed_changed}.
The initial state of the policeman can thus be set such that 
the interrupt happens at any desired point in time.
This is in essence the elapsed time of the atomic model: 
how long it has been since the last transition.
This is therefore the second problem: the time that was already spent in the 
initial state at the start of a simulation cannot be set.

%% file: 2-devs.tex
\section{Background}
\label{sec:devs}
In this section, the Classic DEVS formalism is briefly presented to provide 
the necessary background for our extension to the formalism.
The running traffic light example is subsequently specified in DEVS.
Only Classic DEVS is presented here as our extension is applicable to both 
Classic and Parallel DEVS. 

For a more detailed explanation of the Classic DEVS formalism, we refer the reader to 
the main DEVS reference work~\cite{ClassicDEVS} (including Parallel DEVS) or our 
Classic DEVS tutorial~\cite{WSC_tutorial}.
Further details of Parallel DEVS can be found in the literature 
as well~\cite{ParallelDEVS,ParallelDEVSAbsSim}.

\subsection{Classic DEVS}
Classic DEVS comprises two types of models: Atomic DEVS models, defining behaviour, 
and Coupled DEVS models, defining structure.

An Atomic DEVS model is the basic building block. Its structure is shown 
in Specification~\ref{spec:formal_adevs}.

\begin{align}
    AM = \langle X, Y, S, \delta_{int}, \delta_{ext}, \lambda, ta\rangle
    \label{spec:formal_adevs}
\end{align}
\begin{align*}
    &X                                                      &\textit{set of input events} \\
    &Y                                                      &\textit{set of output events} \\
    &S                                                      &\textit{set of sequential states} \\
    &\delta_{int} : S \to S                                 &\textit{internal transition function} \\
    &\delta_{ext} : Q \times X \to S                        &\textit{external transition function} \\
    &    Q = \{(s, e) | s \in S, 0 \leq e \leq ta(s)        &\textit{set of total states} \\
	&\lambda : S \to Y \cup \{\phi\}                        &\textit{output function, with $\phi$ the null (nothing happens) event} \\
    &ta : S \to \mathbb{R}^+_{0,+\infty}                    &\textit{time advance} \\
\end{align*}

Intuitively, the behavioural semantics are as follows.
The system enters a sequential state $s \in S$ and schedules an ``internal'' transition to state $\delta_{int}(s)$ after $ta(s)$.
Before undergoing the transition, $\lambda(s)$ is invoked to generate an output event $y \in Y$.
If before the scheduled internal transition occurs, an external input event $x \in X$ is received, the scheduled output generation and subsequent transition do not occur. Instead, an ``external'' transition is made to $\delta_{ext}((s, e), x)$.
Here, $e$ is the elapsed time, the time that has passed in the state $s$ since the last transition, until the event was received.
No output is generated in this case.
Upon arrival in the new state, either through $\delta_{int}$ or $\delta_{ext}$, the algorithm repeats.

A Coupled DEVS model is the structuring concept of DEVS, and allows various Atomic and Coupled DEVS models to be combined through parallel composition.
Its structure is shown in Specification~\ref{spec:formal_cdevs}.

\begin{align}
    CM = \langle X_{\textit{self}}, Y_{\textit{self}}, D, MS, IS, ZS, \textit{select} \rangle
    \label{spec:formal_cdevs}
\end{align}
\begin{align*}
    &X_{\textit{self}}                                      &\textit{set of input events} \\
    &Y_{\textit{self}}                                      &\textit{set of output events} \\
    &D                                                      &\textit{set of model instance labels} \\
    &MS = \{M_i | i \in D\}                                 &\textit{set of submodel specifications} \\
	&M_i = \{\langle X_i, Y_i, S_i, \delta_{int, i}, \delta_{ext, i}, \lambda_i, ta_i \rangle | i \in D\} &\textit{(atomic) submodel specification} \\
	&IS = \{I_i | i \in D \cup \{\textit{self}\}\}          &\textit{influencee mapping (encoding connection topology)} \\
    &I_i : i \to 2^{D \cup \{\textit{self}\} \setminus \{i\}} &\textit{set of influencees' labels of model with label i} \\
    &ZS = \{Z_{i,j} | i \in D \cup \{\textit{self}\}, j \in I_i\} &\textit{translation mapping} \\
    &Z_{\textit{self},j} : X_\textit{self} \to X_j          &\textit{input-to-input translation} \\
    &Z_{i,j} : Y_i \to X_j                                  &\textit{output-to-input translation} \\
    &Z_{i,\textit{self}} : Y_i \to Y_\textit{self}          &\textit{output-to-output translation} \\
    &\textit{select} : 2^D \to D                            &\textit{select function}\\
\end{align*}

Intuitively, the semantics are given as follows.
The coupled DEVS model instantiates all of its submodels, which are all considered to be atomic DEVS models (as a coupled model can always be flattened to an atomic model), and keeps their references (or labels) in $D$.
The atomic DEVS submodels' specifications are found in $MS$.
Submodels can be connected and the connection topology is encoded in the 
influencee set $IS$, which lists for each subcomponent, all the other subcomponents 
that it influences (i.e., sends its output to).
Upon forwarding an event to another subcomponent, the event is translated by $ZS$, 
which can map input-to-input (for external input coupling), 
output-to-input (for internal coupling), 
and output-to-output (for external output coupling).
When multiple internal events are scheduled at the same time, in different sub-models, the $\textit{select}$ function is invoked for tie-breaking.

\subsubsection{Running Example}
The Classic DEVS specification of the full traffic light model is given below.
Note that, as per the DEVS semantics, the output function $\lambda$ is invoked \emph{before} the internal transition $\delta_{int}$ is taken. This explains why the produced events are non-intuitive (e.g., raise \EYELLOW for \SGREEN).
The traffic light atomic DEVS model is shown in Specification~\ref{spec:adevs_light}.

\begin{align}
    \textit{Light} = \langle X_{\textit{light}}, Y_{\textit{light}}, S_{\textit{light}}, \delta_{\textit{int}, \textit{light}}, \delta_{\textit{ext}, \textit{light}}, \lambda_{\textit{light}}, \textit{ta}_{\textit{light}} \rangle
    \label{spec:adevs_light}
\end{align}
\begin{align*}
X_{\textit{light}}               = \{&\EAUTO, \EMANUAL\} \\
Y_{\textit{light}}               = \{&\EGREEN, \EYELLOW, \ERED, \EOFF\} \\
S_{\textit{light}}               = \{&\SGREEN, \SYELLOW, \SRED, \STOMANUAL, \STOAUTO, \SMANUAL\} \\
\delta_{\textit{int}, \textit{light}}    = \{&\SGREEN \to \SYELLOW, \SYELLOW \to \SRED, \SRED \to \SGREEN, \\
                            &\STOMANUAL \to \SMANUAL, \STOAUTO \to \SRED\} \\
\delta_{\textit{ext}, \textit{light}}    = \{&(\SGREEN, \_, \EMANUAL) \to \STOMANUAL, (\SYELLOW, \_, \EMANUAL) \to \STOMANUAL, \\
                            &(\SRED, \_, \EMANUAL) \to \STOMANUAL, (\SMANUAL, \_, \EAUTO) \to \STOAUTO\} \\
\lambda_{\textit{light}}         = \{&\SGREEN \to \EYELLOW, \SYELLOW \to \ERED, \\
                            &\SRED \to \EGREEN, \STOMANUAL \to \EOFF, \STOAUTO \to \ERED\} \\
\textit{ta}_{\textit{light}} = \{&\SGREEN \to \DELAYGREEN, \SYELLOW \to \DELAYYELLOW, \SRED \to \DELAYRED, \\
                                 &\SMANUAL \to +\infty, \STOMANUAL \to 0, \STOAUTO \to 0\}
\end{align*}

The policeman's behaviour is simpler and is shown in Specification~\ref{spec:adevs_police}.

\begin{align}
    Police = \langle X_{police}, Y_{police}, S_{police}, \delta_{int, police}, \delta_{ext, police}, \lambda_{police}, ta_{police} \rangle
    \label{spec:adevs_police}
\end{align}
\begin{align*}
X_{police}               &= \{\} \\
Y_{police}               &= \{\EAUTO, \EMANUAL\} \\
S_{police}               &= \{\SBREAK, \SWORKING\} \\
\delta_{int, police}     &= \{\SBREAK \to \SWORKING, \SWORKING \to \SBREAK\} \\
\delta_{ext, police}     &= \{\} \\
\lambda_{police}         &= \{\SBREAK \to \EMANUAL, \SWORKING \to \EAUTO, \\
ta_{police}              &= \{\SBREAK \to \DELAYBREAK, \SWORKING \to \DELAYWORKING\} \\
\end{align*}

Finally, the atomic models are composed in a coupled DEVS model, shown in Specification~\ref{spec:cdevs}.

\begin{align}
    System = \langle X_{\textit{self}}, Y_{\textit{self}}, D, MS, IS, ZS, select \rangle
    \label{spec:cdevs}
\end{align}
\begin{align*}
X_{\textit{self}} = \{&\EAUTO, \EMANUAL\} \\
Y_{\textit{self}} = \{&\EGREEN, \EYELLOW, \ERED, \EOFF\} \\
D = \{&light1, police1\} \\
	MS = \{&M_{light1}=Light, M_{police1}=Police\} \\
IS = \{&light1 \to \{self\}, self \to \{light1\}, police1 \to \{light1\}\} \\
ZS = \{&Z_{\textit{self}, \textit{light1}} = \{\EAUTO \to \EAUTO, \EMANUAL \to \EMANUAL\}, \\
       &Z_{\textit{police1}, \textit{light1}} = \{\EAUTO \to \EAUTO, \EMANUAL \to \EMANUAL\}, \\
       &Z_{\textit{light1}, \textit{self}} = \{\EGREEN \to \EGREEN, \EYELLOW \to \EYELLOW, \\
       &                                       \hspace{2cm}\ERED \to \ERED, \EOFF \to \EOFF\}\} \\
select = \{&\{light1, police1\} \to police1, \{ligth1\} \to light1, \{police1\} \to police1\} \\
\end{align*}

From this complete DEVS specification example, it is clear that something is lacking: nowhere has been specified what is the initial state, and how long the system has already been in that state.

%% file: 3-initial_total_state.tex
\section{Initial Total State}
\label{sec:total_state}
After the above brief recap of the Classic DEVS formalism and the presentation of our running example in this specification, it has become clear that this model cannot be simulated as-is.
Indeed, a simulator for this model has no way of knowing from which total state to start the simulation.
To alleviate this problem, we propose to add an \emph{initial total state} to the DEVS specification.
While a minimal extension, this has repercussions on both syntax and semantics of the DEVS formalism.

\subsection{Atomic DEVS Specification}
In the Atomic DEVS specification, the total initial state $q_{\textit{init}} \in Q$ is added.
$Q$ was previously described as being the set of total states $\{(s, e) | s \in S, 0 \leq e \leq \textit{ta}(s)\}$, as used by the external transition function $\delta_{\textit{ext}}$.
This alters the atomic DEVS specification to the following.
\begin{align*}
    AM = \langle X, Y, S, q_{\textit{init}}, \delta_{\textit{int}}, \delta_{\textit{ext}}, \lambda, \textit{ta} \rangle
\end{align*}

The initial total state comprises the initial state $s_{\textit{init}}$ and the initial elepased time $e_{\textit{init}}$.

The initial state $s_{\textit{init}} \in S$ specifies the system state in which the simulation commences.
Its addition is logical, and has up to now been implemented in different simulation tools, as an implicit ad-hoc extension to the formalism.
In the case of our running example, we may specify that the traffic light starts in the \SGREEN state and the policeman in the \SBREAK state.

The initial elapsed time $e_{\textit{init}}$ specifies how long the system has been in this state, without a transition being observed.
This is a less obvious addition to the initialization phase, and is ignored by several simulation tools.
Nonetheless, we argue for its importance in providing flexibility to the DEVS modeller.

If only an $s_{\textit{init}}$ is present in the specification, but not $e_{\textit{init}}$,
it is possible to specify \SGREEN and \SBREAK as the initial state for the traffic light and policeman, respectively, but one is restricted to their time advances.
Indeed, without $e_{\textit{init}}$, the initial elapsed time would be implicitly equal to $0$, as was shown in Figure~\ref{fig:trace_elapsed_0}.
This schedules the first internal transition of the policeman at time \DELAYBREAK.
The traffic light will have internal transitions at times 57 (to \SYELLOW), 60 (to \SRED), and 120 (to \SGREEN).
Following this sequence, the policeman will \emph{always} send its interrupt at the exact same point in time, namely when a switch is made from \SRED to \SGREEN.
Therefore, it is impossible for the modeller to reproduce the real-world scenario where, for example, the policeman interrupts after the light has been in the \SYELLOW state for $1.5$ time units.
To do this, the model itself would have to be drastically modified (e.g., adding an artificial ``initialization'' state to the policeman model). This severely impedes modular re-use of submodels.

\begin{figure}
    \center
    \includegraphics[width=0.60\textwidth]{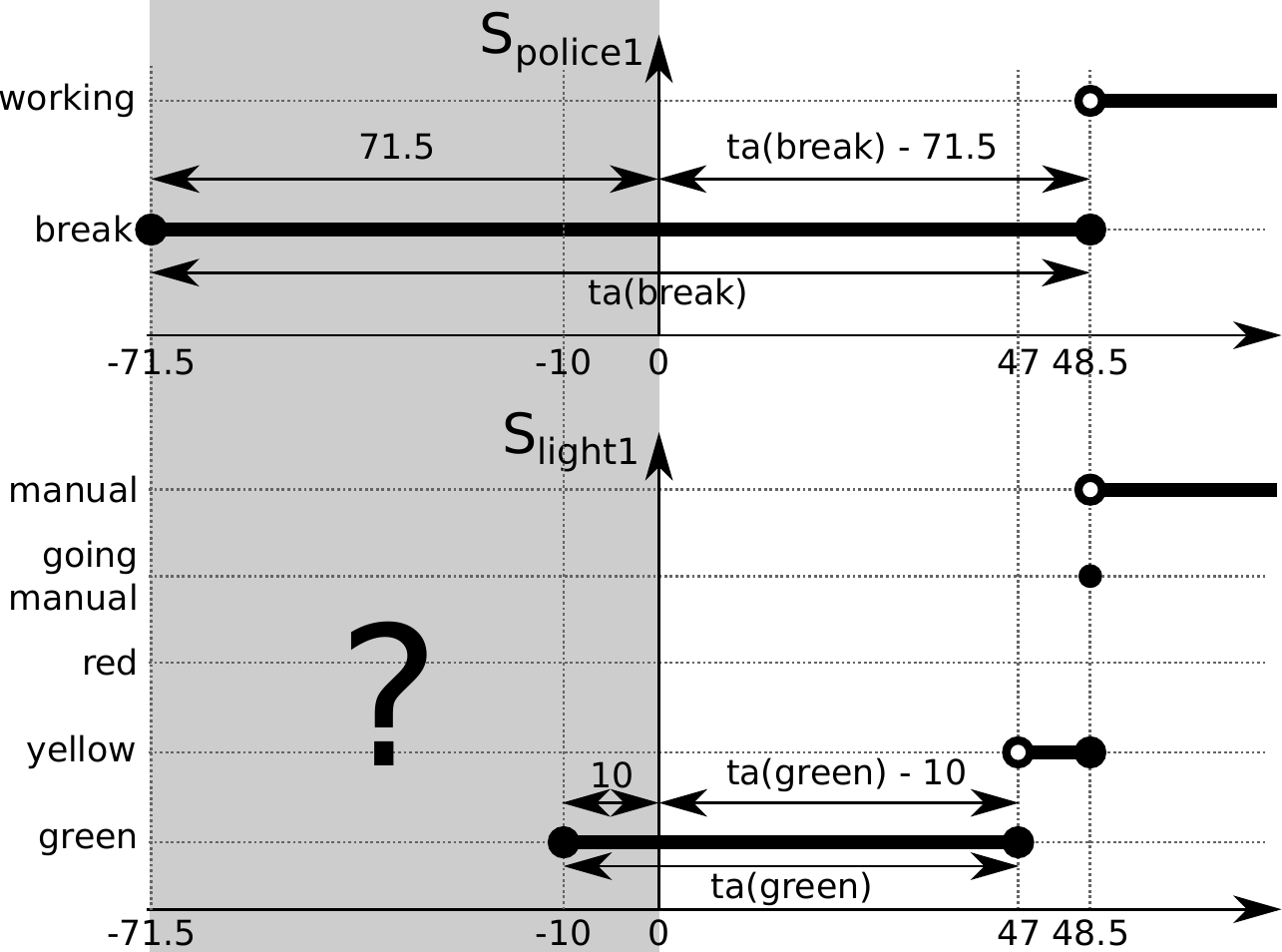}
    \caption{Simulation trace including hypothetical negative simulation time (grayed out).}
    \label{fig:trace_elapsed_required}
\end{figure}

\begin{figure}
    \center
    \includegraphics[width=0.60\textwidth]{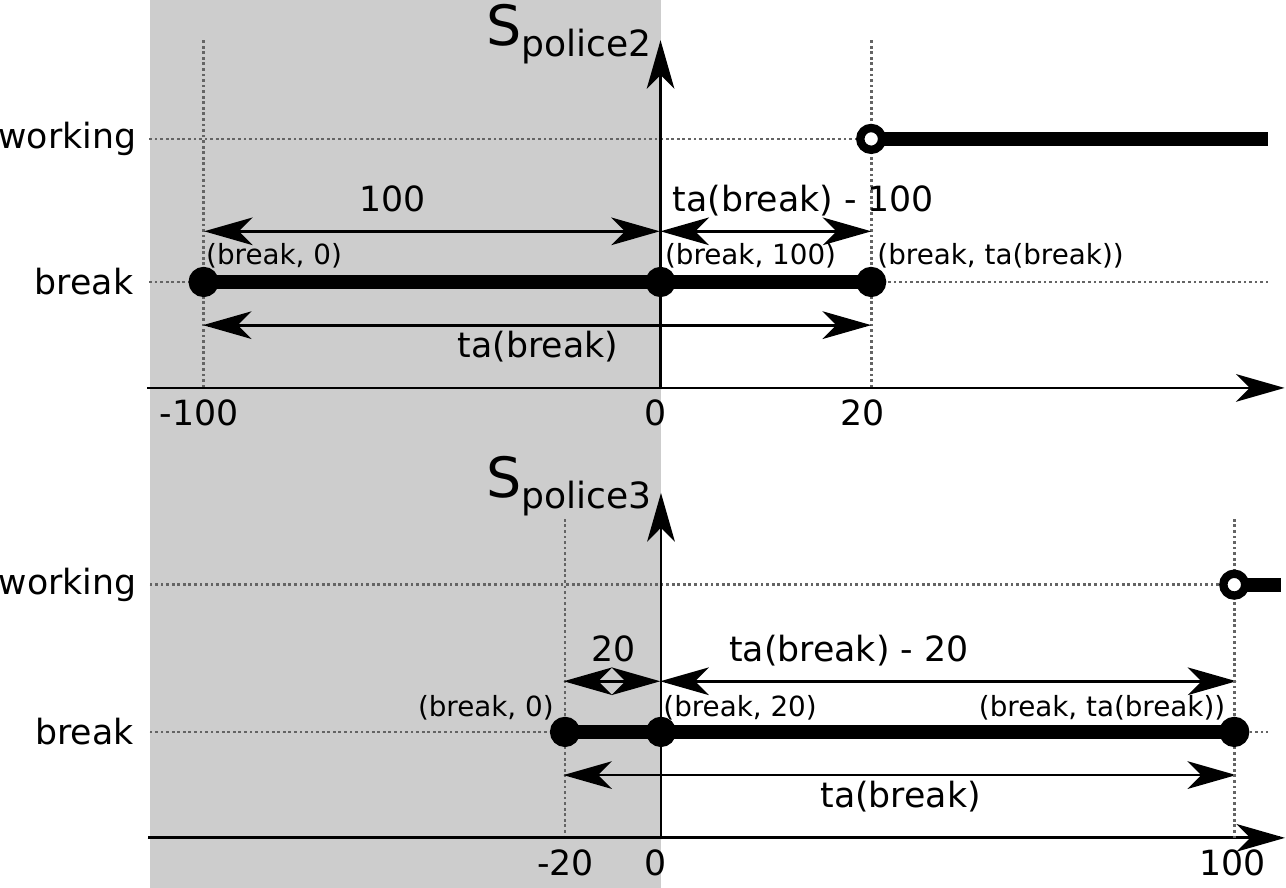}
    \caption{Various options for shifting the police model.}
    \label{fig:trace_elapsed_various}
\end{figure}

What we wish to achieve is shown in Figure~\ref{fig:trace_elapsed_required}, which includes the ``negative'' simulation time (i.e., what hypothetically happened before simulation, given the specified model).
While this figure includes the state trace from before the start of the simulation, we can only go back until the last transition, as we have no knowledge about how we ended up in that state (e.g., before time $-10$ for \textit{police1}).
To remain consistent with the DEVS specification, it is required to alter the duration since the last event for each atomic model individually.
By doing this, each individual atomic DEVS model can be shifted relatively to all others.
For example, Figure~\ref{fig:trace_elapsed_various} presents two different initial elapsed time configurations, with their effect on the simulation.
Note that these additions can easily be taken over to Parallel DEVS as well, without any changes.

In our running example, we augment the models with the following initial total states.
For the sake of the closure under coupling, we set the initial elapsed time of the traffic light to 10.
Therefore, the light will enter \SYELLOW at time 47.
To ensure that the policeman sends the external interrupt after the light has been in state \SYELLOW for $1.5$, the transition has to happen at time $48.5$.
To achieve this, we compute its desired initial elapsed time as $e_{\textit{init}, \textit{police1}} = \DELAYBREAK - 48.5 = 71.5$.
Thus, the policeman will have already spent $71.5$ time units in \SBREAK, and will transition after $48.5$ time units, at which point the light will have been in \SYELLOW for $1.5$ time units, as desired.
Various configurations exist to achieve the same result, such as having the traffic light start at a different state or with a different elapsed time.
\begin{align*}
    q_{\textit{init}, \textit{light1}} = &(\SGREEN, 10) & q_{\textit{init}, \textit{police1}} = &(\SBREAK, 71.5)
\end{align*}

\subsection{Closure Under Coupling}
The atomic DEVS specification extension has its repercussions on closure under coupling, where several atomic DEVS models are flattened to a single atomic DEVS model.
Indeed, the $q_{\textit{init}}$ of each submodel, has to be conserved in the flattened atomic model.
While several approaches exist, we have opted for a non-invasive approach, which only affects $q_{\textit{init}}$.

A definition for $q_{\textit{init}}$ in terms of the $q_{\textit{init}, i}$ ($i$ being a submodel label) is presented which leaves the closure under coupling untouched.
As $q_{\textit{init}} \in Q$, it consists of two parts: the initial state $s_{\textit{init}}$ and the initial elapsed time $e_{\textit{init}}$.

First, we tackle the initial elapsed time $e_{\textit{init}}$.
As closure under coupling combines all subcomponents, every independent state change of a subcomponent results in a state change of the flattened model.
As a logical consequence, the initial elapsed time of the overall system is the time since the last transition of all its subcomponents, i.e., the minimum of all the subcomponents' elapsed times.
\[
    e_{\textit{init}} = min_{i \in D} \{e_{\textit{init}, i}\}
\]

\begin{figure}
    \center
    \includegraphics[width=0.80\textwidth]{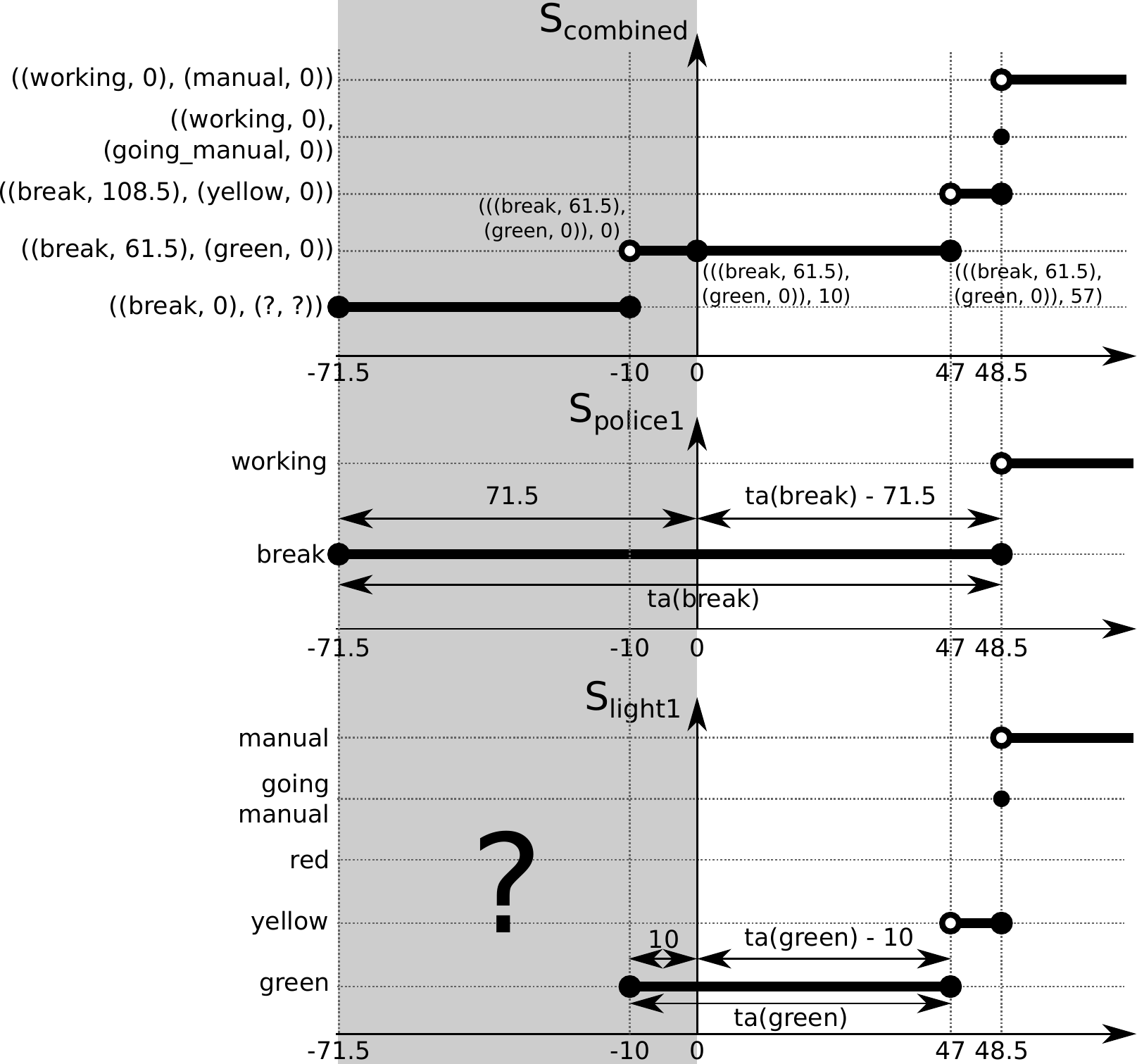}
    \caption{$q_{\textit{init}}$ in the flattened model.}
    \label{fig:closure_elapsed}
\end{figure}

Second, we tackle the initial state $s_{\textit{init}}$.
As mentioned earlier, closure under coupling combines the states of the various subcomponents into a single state.
To capture the individual elapsed times, required for the $\delta_{\textit{ext}}$ and $\delta_{\textit{int}}$ functions, the flattened atomic DEVS model not only encodes the combination of all subcomponents' states, but includes for each subcomponent, the \emph{current elapsed time}: $S$ is defined as $\times_{i \in D} Q_i$.
Intuitively, one would define $s_{\textit{init}}$ as the combination of the $q_{\textit{init}, i}$ values of all the subcomponents, i.e.,  $s_{\textit{init}} = \times_{i \in D} q_{init}$, but this is incorrect.
Due to the definition of $e_{\textit{init}}$, the minimal elapsed time is already taken into account, as shown in Figure~\ref{fig:closure_elapsed}.
Indeed, it has been $e_{\textit{init}}$ time units since the last transition (to $s_{\textit{init}}$), and therefore $s_{\textit{init}}$ should not contain the elapsed times at initialization, but rather those upon reaching the initial state (i.e., $e_{\textit{init}}$ time units ago).
This can be done by subtracting from each $e_{\textit{init}, i}$ in $s_{\textit{init}}$, the globally minimum value $e_{\textit{init}}$.
Therefore, we define $s_{\textit{init}}$ as follows.
\[
    s_{\textit{init}} = (..., (s_{\textit{init}, i}, e_{\textit{init}, i} - e_{\textit{init}}), ...)
\]

An alternative would be to set $e_{\textit{init}}$ to 0 and define $s_{\textit{init}}$ as the combination of all $q_{\textit{init}, i}$ values, but this is less intuitive for two reasons.
First, the combination of $q_{\textit{init}, i}$ is never actually ``materialized'' in simulators, as discrete event simulators jump from one transition time to the next, ignoring all intermediate times (this is the essence of discrete-event simulation).
In many cases, no transition happens at simulation time zero, and therefore the state would never have materialized anyway.
The option of altering this behaviour for the initialization seems non-intuitive.
Second, to keep the definition of elapsed time consistent with that used by the simulator ``in regime operation'', it must be the time since the last transition of the atomic DEVS model.
Setting it to 0 it implies that the system made a transition upon initialization, which is incorrect, unless of course in some rare cases where this occurs in the system being modelled.

For our running example, using the same parameters as before, this results in the following initial total state of the flattened atomic DEVS model.
\[
    q_{init} = (((\SGREEN, 0), (\SBREAK, 61.5)), 10)
\]

Note that these additions can easily be taken over to Parallel DEVS as well, without any changes.

\subsection{Abstract Simulator}
The abstract simulator is simpler to alter, as there is already a notion of \emph{initialization message}, often termed $i$.
At the start of the simulation, the root coordinator first sends out the message $(i, 0.0)$.
Each coordinator or simulator responds to this message with $(done, tn)$, where $tn$ represents the earliest internal event it has scheduled.
As in our extended DEVS specification, the atomic models already have the necessary initialization information stored locally, they can set the state and elapsed time upon receiving the $i$ message.
Depending on these newly set values, the initial $ta(s)$ is computed, which is then sent to the parent simulator.
No further changes are required.

%% file: 5-related_work.tex
\section{Related Work}
\label{sec:related_work}
The initialization of DEVS models has up to now been done either by the experimental frame~\cite{ClassicDEVS,muzy_experimental_frame}, or hard coded in tools.

In the experimental frame approach, the initial total state can be considered as a type of input.
The initial total state is thus defined in the experimental frame, and is merely passed along to the DEVS model during initialization.
While this approach allows the environment to configure the model as desired, thereby offering flexibility, it is not formalized in the DEVS specification.
Indeed, the DEVS semantics, whether it be denotationally using closure under coupling or operationally using the abstract simulator, don't mention anything related to the initial total state.
Most of the time, it is a logical decision to have the model itself define what is a consistent state for initialization.

Initialization is often left to tools, which implement this independent of the formalim.
We consider several popular DEVS simulation tools, discussed in an earlier survey~\cite{DEVSSurvey}.

\textbf{Adevs}~\cite{adevs} considers all attributes of an atomic model to be part of the state. Therefore, the constructor is responsible for setting all attributes.
The initial elapsed time cannot be set.

\textbf{CD++}~\cite{cd++} similarly considers all attributes to be part of the state, but provides a specific function called \texttt{initFunction}.
This function is called upon initialization and sets the initial state and the remaining time in this state ($\sigma$).
While this is functionally equivalent to the elapsed time, as $e = ta(s) - \sigma$, it diverges from the DEVS specification, as it is possible to be in an initial state which would have ordinarily been impossible to reach (e.g., $q_{init} = (\SYELLOW, 100)$ for our running example).
Indeed, it might be that $\sigma > ta(s)$, which will remain undetected.
In contrast, if the elapsed time is given, the definition of $Q$ ensures that $0 \leq e \leq ta(s)$, rendering such inconsistencies impossible.

\textbf{DEVS-Suite}~\cite{DEVS-Suite} similarly has an \texttt{initialize} function which initializes the state and sets the first timeout to use, similar to CD++.
As its approach is the same as that of CD++, it shares the same problems.

\textbf{MS4Me}~\cite{MS4Me} again uses the same approach, but requires the syntax \texttt{to start hold in S\_INIT for time REMAINING\_INIT!} to achieve the same result.
Again, the same problems occur.

\textbf{PowerDEVS}~\cite{PowerDEVS} has the same approach, where the function is called \texttt{init}.
Again, the function initializes the state and returns the remaining time until the first internal transition.

\textbf{PythonPDEVS}~\cite{PythonPDEVS1}, our Classic and Parallel DEVS simulator, supports the definition of an initial total state as presented in this paper.
Like all tools, it is required to specify the initial state (assign to \texttt{self.state}), but additionally, the elapsed time can be set (assign to \texttt{self.elapsed}).
By specifying the elapsed time, instead of the remaining time, the timings are guaranteed to be consistent.

\textbf{VLE}~\cite{vle} uses the constructor to initialize the state, as in adevs, and similarly does not present an option to specify the initial elapsed or remaining time.

\textbf{X-S-Y}~\cite{X-S-Y} assigns the initial state to the \texttt{self.phase} attribute, but does not allow for either the elapsed or remaining time to be set.

While several tools have addressed the missing initialization in the DEVS specification, they do so in widely varying ways.
Syntactical differences are found among all tools, as there is no agreed upon name or method of adding initialization information.
More importantly, semantical differences exist as well: tools such as adevs, vle, and X-S-Y can not specify the initial time, and are therefore restricted in their (model reuse) flexibility.
Other tools, such as CD++, DEVS-Suite, MS4Me, and PowerDEVS, allow the remaining time to be specified, though this potentially results in inconsistent situations, as there is no formal constraint on this value.
Finally, PythonPDEVS implements the initial total state using the elapsed time, thereby performing consistency checking: the elapsed time $e$ is immediately compared to the time advance $ta$.
In the light of DEVS standardization efforts~\cite{standardDEVS1,standardDEVS2}, it is problematic that various tools implement different methods of initialization.
On this topic, a ``DEVS compliance checklist'' was previously introduced~\cite{compliance}, and later slightly extended~\cite{DEVSSurvey}, to which we believe the initial total state should now be added.

%% file: 6-conclusion.tex
\section{Conclusion}
\label{sec:conclusion}
DEVS has since long been acclaimed for its rigourous and precise specification for models with a discrete event abstraction.
Despite its rigour being one of its main advantages, model initialization is left unspecified, leading to various implementations.
This paper has presented an addition to the DEVS specification, the initial total state $q_{init}$, which formalizes model initialization.
The influence of this addition on the specification and closure under coupling was presented in detail, both at the level of the specification, and using a traffic light example. 
Classic DEVS was used to present our contribution, though the addition can be applied to Parallel DEVS as well.